\documentclass[aps,prl,preprint,tightenlines,superscriptaddress,showpacs,byrevtex]{revtex4}

\usepackage{graphicx} 
\usepackage{dcolumn}  

\graphicspath{{ps}}

\newcommand{\gev}{\ensuremath{\mathrm{GeV}}}
\newcommand{\like}{\ensuremath{\mathcal{L}}}
\newcommand{\lrat}{\ensuremath{\mathcal{P}}}

\begin{document}



\title{ \quad\\[0.5cm]  Measurement of masses  of the~$\Xi_c(2645)$ 
and~$\Xi_c(2815)$ baryons 
and observation of  $\Xi_c(2980)\to \Xi_c(2645)\pi$}



\affiliation{Budker Institute of Nuclear Physics, Novosibirsk, Russia}
\affiliation{Chiba University, Chiba, Japan}
\affiliation{Justus-Liebig-Universit\"at Gie\ss{}en, Gie\ss{}en, Germany}
\affiliation{The Graduate University for Advanced Studies, Hayama, Japan}
\affiliation{Gyeongsang National University, Chinju, South Korea}
\affiliation{Hanyang University, Seoul, South Korea}
\affiliation{University of Hawaii, Honolulu, HI, USA}
\affiliation{High Energy Accelerator Research Organization (KEK), Tsukuba, Japan}
\affiliation{Hiroshima Institute of Technology, Hiroshima, Japan}
\affiliation{Institute of High Energy Physics, Chinese Academy of Sciences, Beijing, PR China}
\affiliation{Institute for High Energy Physics, Protvino, Russia}
\affiliation{Institute of High Energy Physics, Vienna, Austria}
\affiliation{Institute for Theoretical and Experimental Physics, Moscow, Russia}
\affiliation{J. Stefan Institute, Ljubljana, Slovenia}
\affiliation{Kanagawa University, Yokohama, Japan}
\affiliation{Korea University, Seoul, South Korea}
\affiliation{Kyungpook National University, Taegu, South Korea}
\affiliation{\'Ecole Polytechnique F\'ed\'erale de Lausanne, EPFL, Lausanne, Switzerland}
\affiliation{Faculty of Mathematics and Physics, University of Ljubljana, Ljubljana, Slovenia}
\affiliation{University of Maribor, Maribor, Slovenia}
\affiliation{University of Melbourne, Victoria, Australia}
\affiliation{Nagoya University, Nagoya, Japan}
\affiliation{Nara Women's University, Nara, Japan}
\affiliation{National Central University, Chung-li, Taiwan}
\affiliation{National United University, Miao Li, Taiwan}
\affiliation{Department of Physics, National Taiwan University, Taipei, Taiwan}
\affiliation{H. Niewodniczanski Institute of Nuclear Physics, Krakow, Poland}
\affiliation{Nippon Dental University, Niigata, Japan}
\affiliation{Niigata University, Niigata, Japan}
\affiliation{University of Nova Gorica, Nova Gorica, Slovenia}
\affiliation{Osaka City University, Osaka, Japan}
\affiliation{Osaka University, Osaka, Japan}
\affiliation{Panjab University, Chandigarh, India}
\affiliation{Saga University, Saga, Japan}
\affiliation{University of Science and Technology of China, Hefei, PR China}
\affiliation{Seoul National University, Seoul, South Korea}
\affiliation{Sungkyunkwan University, Suwon, South Korea}
\affiliation{University of Sydney, Sydney, NSW, Australia}
\affiliation{Tata Institute of Fundamental Research, Mumbai, India}
\affiliation{Toho University, Funabashi, Japan}
\affiliation{Tohoku Gakuin University, Tagajo, Japan}
\affiliation{Tohoku University, Sendai, Japan}
\affiliation{Department of Physics, University of Tokyo, Tokyo, Japan}
\affiliation{Tokyo Institute of Technology, Tokyo, Japan}

\affiliation{Tokyo Metropolitan University, Tokyo, Japan}

\affiliation{Tokyo University of Agriculture and Technology, Tokyo, Japan}
\affiliation{Virginia Polytechnic Institute and State University, Blacksburg, VA, USA}
\affiliation{Yonsei University, Seoul, South Korea}

\author{T.~Lesiak}\affiliation{H. Niewodniczanski Institute of Nuclear Physics, Krakow, Poland} 
  \author{I.~Adachi}\affiliation{High Energy Accelerator Research Organization (KEK), Tsukuba, Japan} 
  \author{H.~Aihara}\affiliation{Department of Physics, University of Tokyo, Tokyo, Japan} 
  \author{K.~Arinstein}\affiliation{Budker Institute of Nuclear Physics, Novosibirsk, Russia} 
  \author{A.~M.~Bakich}\affiliation{University of Sydney, Sydney, NSW, Australia}
  \author{V.~Balagura}\affiliation{Institute for Theoretical and Experimental Physics, Moscow, Russia} 
  \author{E.~Barberio}\affiliation{University of Melbourne, Victoria, Australia} 
  \author{I.~Bedny}\affiliation{Budker Institute of Nuclear Physics, Novosibirsk, Russia} 
  \author{K.~Belous}\affiliation{Institute for High Energy Physics, Protvino, Russia} 
  \author{V.~Bhardwaj}\affiliation{Panjab University, Chandigarh, India} 
% ...

\collaboration{The Belle Collaboration}

\begin{abstract}
We report a precise measurement of the masses of the $\Xi_c(2645)$ and 
$\Xi_c(2815)$ baryons
using a data sample of 414~fb$^{-1}$ collected by the Belle collaboration
at the KEKB $e^+ e^-$ collider.
The states $\Xi_c(2645)^{0,+}$ 
are observed in the  $\Xi_c^{+,0}\pi^{-,+}$ decay modes,
while the $\Xi_c(2815)^{0,+}$
are reconstructed in the $\Xi_c(2645)^{+,0}\pi^{-,+}$ decay modes.
The following mass splittings are determined:
$m_{\Xi_c(2645)^+} - m_{\Xi_c(2645)^0} = (-0.1 \pm 0.3 ({\rm stat}) \pm 0.6 {(\rm syst}))~{\rm MeV}/{\rm c}^2$ and
$m_{\Xi_c(2815)^+} - m_{\Xi_c(2815)^0} =  (-3.4 \pm 1.9 ({\rm stat}) \pm 0.9 {(\rm syst}))~{\rm MeV}/{\rm c}^2$
with a much better precision than the current world averages.
We also observe a new decay mode, $\Xi_c(2980)^{0,+}  \to  \Xi_c(2645)^{+,0}\pi^{-,+}$. 
\end{abstract}


\maketitle

\tighten


\section{Introduction}


The study of charmed baryons has recently been a focus of
significant experimental
effort~\cite{RUSLAN,BABARXC,MIZUK,BABARLC,BABARD0,BABAROM,CLEOSIG,XC1}. 
Several new excited
states, such as the $\Xi_c(2980)$, $\Xi_c(3055)$, $\Xi_c(3077)$, $\Xi_c(3123)$
 and $\Lambda_c(2940)$ 
have been observed, or their properties determined for the first
time, enabling tests of quark (and other) models and predictions of
heavy quark symmetry~\cite{HQS1,HQS2}.

This paper presents a study of exclusive decays of 
the $\Xi_c(2645)^0$, $\Xi_c(2645)^+$, $\Xi_c(2815)^0$ and $\Xi_c(2815)^+$ 
 baryons~\cite{CHGCONJ}
 and a determination of their masses and the corresponding mass splittings within
isospin doublets.  The $\Xi_c(2645)^0$ and
$\Xi_c(2645)^+$ are reconstructed in the $\Xi_c^+\pi^-$ and $\Xi_c^0\pi^+$
decay modes, respectively. The latter mode was first observed by the
CLEO collaboration~\cite{CLEO1}, while the former decay mode is observed here for the first time.
  For the hyperons $\Xi_c(2815)^0$ and
$\Xi_c(2815)^+$, first seen by the CLEO collaboration~\cite{CLEO2},
the decays into
$\Xi_c(2645)^+\pi^-$ and $\Xi_c(2645)^0\pi^+$ are observed.

In the mass spectra of  $\Xi_c(2645)^{+,0}\pi^{-,+}$ pairs, we observe clear peaks 
close to the  $\Xi_c(2980)^{0,+}$ reported by the Belle~\cite{RUSLAN} 
and BaBar~\cite{BABARXC} collaborations in the
$\Lambda_c^+ K^- \pi^+$ and $\Lambda_c^+ K^0_S\pi^-$ final states.

This article is organized as follows. In the  first two sections we describe
the data sample and the reconstruction of~$\Xi_c$ baryons.
The next two sections are devoted to the precise determination of the
$\Xi_c(2645)$ and $\Xi_c(2815)$ masses. Finally, in the last section,
we discuss the mass peaks observed 
above the $\Xi_c(2815)^{+,0}$ states in the $\Xi_c(2645)^{0,+}\pi^{+,-}$ systems.

 
\section{Detector and data Sample}

The data used for this study were collected on the $\Upsilon(4S)$ resonance
using the Belle detector at the KEKB asymmetric-energy $e^+e^-$ collider~\cite{KEKB}.
The integrated luminosity of the data sample is~414~fb$^{-1}$. 

The Belle detector is a large-solid-angle magnetic spectrometer that consists of 
a~silicon vertex detector (SVD), a 50-layer central drift chamber (CDC),
an array of aerogel threshold Cherenkov counters (ACC),
 a barrel-like arrangement of 
time-of-flight scintillation counters (TOF), and an electromagnetic calorimeter 
comprised of CsI(Tl) crystals (ECL) located inside a superconducting solenoid coil 
that provides a 1.5~T magnetic field. An iron flux-return located outside of the  
coil is instrumented to detect $K^0_L$ mesons and to identify muons (KLM). 
A detailed  description of the Belle detector can be found elsewhere~\cite{BELLE}.


\section{Reconstruction}
\label{RECON}

Reconstruction of $\Xi_c$, $\Xi_c(2645)$ and $\Xi_c(2815)$
 decays for this analysis proceeds in three steps:
reconstruction of tracks and their identification as protons, kaons or pions;
combination of tracks to reconstruct $\Lambda$ and $\Xi^-$
hyperons;
and the selection of $\Xi_c$ candidates from combinations of tracks
and hyperons.
The method used for each step is described in the following sections.


\begin{figure}[tbh]
\begin{minipage}[b]{.46\linewidth}
\centering
\setlength{\unitlength}{1mm}
\begin{picture}(95,85)
\put(65,77){\Large\bf (a)}
\put(-5,35){\rotatebox{90}{\large\bf Events / (2.5~{\rm MeV}/c$^2$)}}
\put(25,-1){{\Large $m(\Xi_c^+\pi^-)$ [GeV/c$^2$]}}
\includegraphics[height=9.5cm,width=8.5cm]{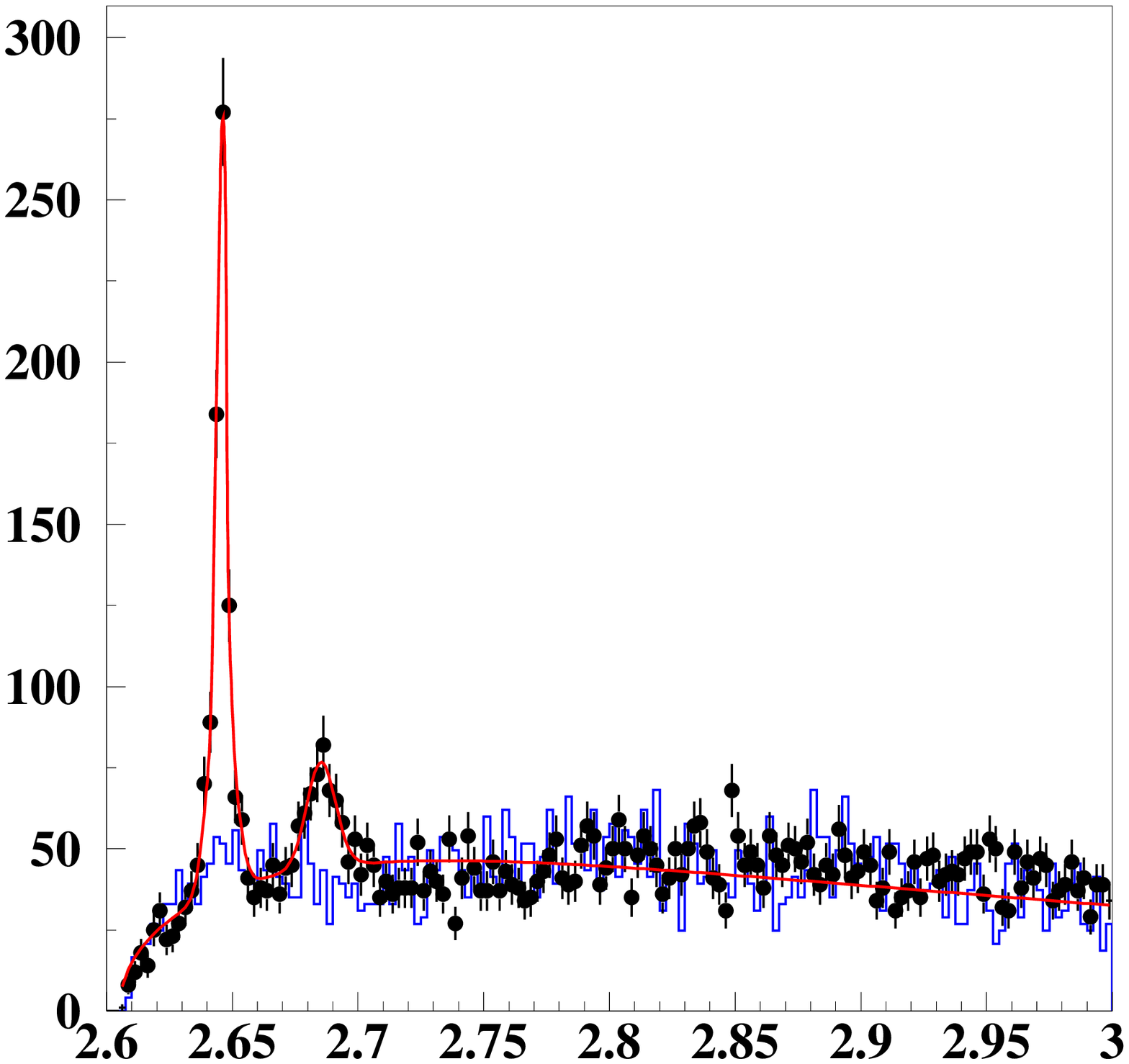}
\end{picture}
\end{minipage}\hfill
\begin{minipage}[b]{.46\linewidth}
\centering
\setlength{\unitlength}{1mm}
\begin{picture}(95,85)
\put(65,77){\Large\bf (b)}
\put(25,-1){{\Large $m(\Xi_c^0\pi^+)$ [GeV/c$^2$]}}
\put(-5,35){\rotatebox{90}{\large\bf Events / (2.5~{\rm MeV}/c$^2$)}}
\includegraphics[height=9.5cm,width=8.5cm]{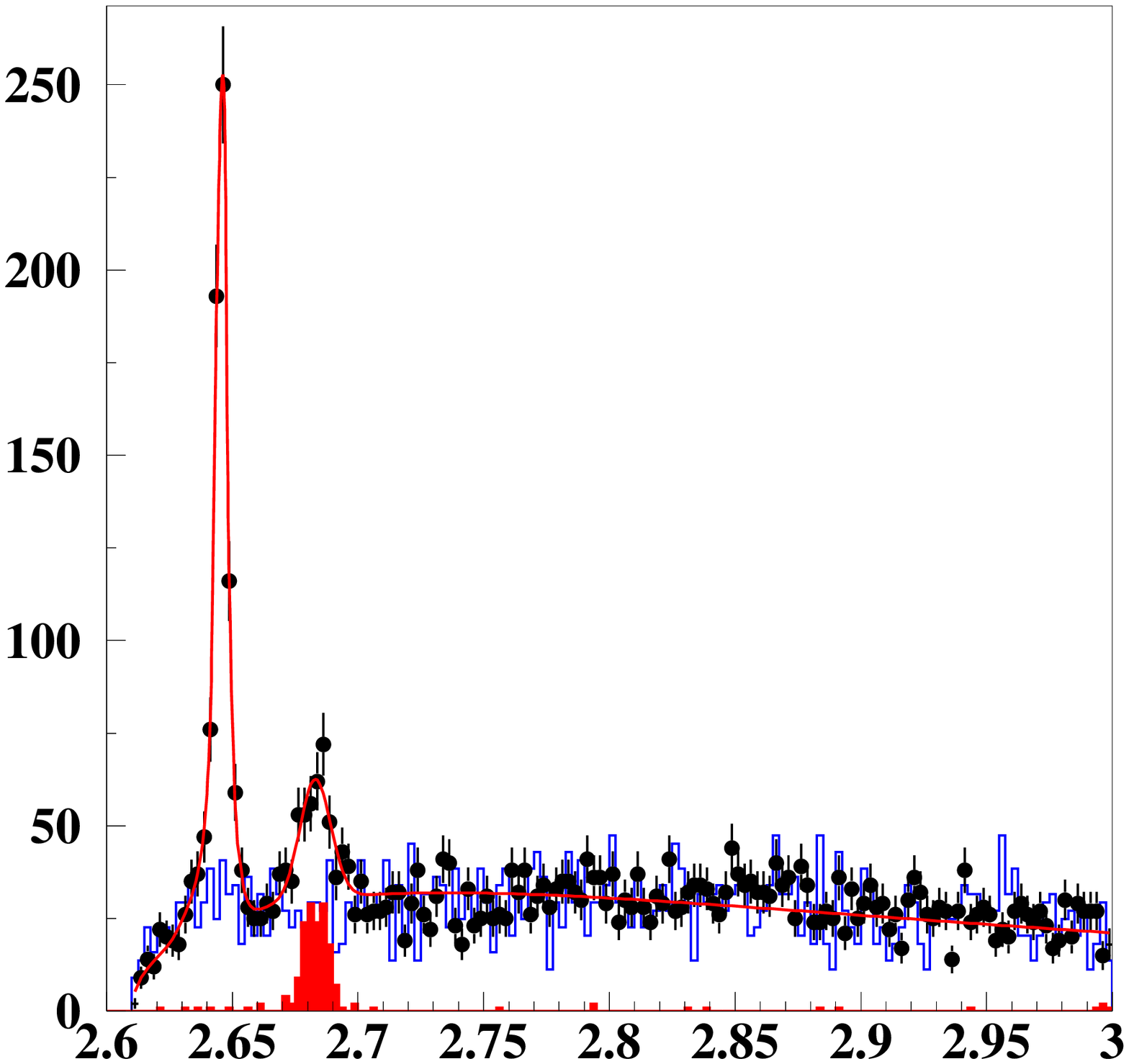}
\end{picture}
\end{minipage}
\caption{Invariant mass distributions for 
{\bf (a)} $\Xi_c^+\pi^-$ ($\Xi_c^+\to\Xi^-\pi^+\pi^+$) and
{\bf (b)} $\Xi_c^0\pi^+$ ($\Xi_c^0\to\Xi^-\pi^+$).
Curves correspond to the fit described in the text. The histograms  
show the $\Xi_c$ mass sidebands. The second peak is due to the feed-down from 
$\Xi_c(2790)\to \Xi_c^{\prime}(2579)\pi, \Xi_c^{\prime}\to\Xi_c\gamma$, as determined
from Monte Carlo simulations and marked in {\bf (b)} as a shaded histogram.} 
\label{FIG_XC2645}
\end{figure}

\subsection{Track reconstruction and identification}

Charged tracks are reconstructed from hits in the CDC using a Kalman 
filter~\cite{KALMAN}
and matched to hits in the SVD.
Quality criteria are then applied.
All tracks other than those used to form $\Lambda$ and $\Xi^-$ candidates,
are required to have impact parameters relative to the
interaction point (IP) of less than 0.5~cm in the $r-\phi$ plane,
and 5~cm in the $z$ direction~\cite{ZAXIS}.
The transverse momentum of each track is required to exceed $0.1\,\gev/c$,
in order to reduce the low momentum combinatorial background.

Hadron identification  is based on information from
the CDC (energy loss $dE/dx$), TOF and ACC, combined to form likelihoods 
$\like(p)$, $\like(K)$ and $\like(\pi)$ for the proton, kaon and
pion hypotheses, respectively.
These likelihoods are combined to form ratios
$\lrat(K/\pi) = \like(K) / (\like(K) + \like(\pi))$
and $\lrat(p/K) = \like(p) / (\like(p) + \like(K))$,
spanning the range from zero to one, which are then used to identify individual tracks~\cite{BELLE}.
Pion candidates, except those coming from the decay of the 
$\Lambda$ hyperon, should satisfy both a proton and a kaon veto: 
$\lrat(p/K) < 0.98$ and $\lrat(K/\pi) < 0.98$. 

Electrons are identified using a similar likelihood ratio
$\lrat_e = \like_e / (\like_e + \like_{\text{non-}e})$,
based on a combination of $dE/dx$ measurements in the CDC, 
the response of the ACC, $E/p$, where $p$ is the momentum of the track
and $E$ the energy of the associated cluster in the ECL, as well as matching between the track and 
the ECL cluster position and the transverse shower shape.  
All tracks with $\lrat_e > 0.98$ are assumed to be electrons,
and removed from the proton, kaon and pion samples.


\subsection{Reconstruction of $\Lambda$ and $\Xi^-$}

We reconstruct $\Lambda$ hyperons in the $\Lambda\to p\pi^-$ decay mode,
requiring the proton track to satisfy $\lrat(p/K) > 0.1$~\cite{PROTON}, and fitting the
$p$ and $\pi$ tracks to a common vertex.
To reduce the number of poorly reconstructed candidates,
the $\chi^2/n.d.f.$\cite{CHI2}
 of the vertex should not exceed 25
 (removing  approximately 2\% of signal candidates)
and the difference in the $z$-coordinate between the proton and pion at the 
vertex is required to be less than 2 cm. Due to the large $c\tau$ factor for $\Lambda$
hyperons (7.89~cm), we demand that the distance between the decay vertex and 
the IP in the $r-\phi$ plane be greater than 1~cm.
The invariant mass of the proton-pion pair is required to be within
2.4 MeV/c$^2$ ($\approx 2.5$ standard deviations) of the nominal $\Lambda$ mass.
The mean  value of the  $\Lambda$ signal in the reconstructed mass distribution 
is found to be $1115.7\pm 0.1$ MeV/c$^2$, in agreement with the world average value~\cite{PDG}.

We reconstruct $\Xi^-$ hyperons in the decay mode $\Xi^-\to \Lambda\pi^-$.
The $\Lambda$ and~$\pi$ candidates are fitted to a common vertex, 
for which we require  $\chi^2/n.d.f. < 25$
 (removing approximately  2\% of signal candidates).
The distance between the $\Xi^-$ decay vertex position and the IP in the 
$r-\phi$ plane should be at least 5~mm, and
less than the corresponding distance between the IP and the $\Lambda$ vertex.
The invariant mass of the $\Lambda \pi^-$ pair is required to be within
7.5 MeV/c$^2$ of the nominal value ($\approx 2.5$ standard deviations).
The mass of the $\Xi^-$ is found to be $1321.78\pm 0.21$ MeV/c$^2$, in agreement with
the PDG average: $1321.34\pm 0.14$ MeV/c$^2$~\cite{PDG}.


\subsection{Reconstruction of $\Xi_c$, $\Xi_c(2645)$ and $\Xi_c(2815)$}


The reconstructed $\Lambda$ and $\Xi^-$ candidates and the remaining
charged hadrons in an event
are combined to form candidates for
the  decays $\Xi_c^+  \to \Xi^-\pi^+\pi^+$ 
and $\Xi_c^0  \to  \Xi^-\pi^+$. 
The signal region  is defined by the reconstructed
mass windows 
(2.455--2.485)~GeV/c$^2$ for the former, and 
(2.45--2.49)~GeV/c$^2$ for the latter decay.
All particles forming the $\Xi_c$ candidate are then fitted to a common vertex
constraining their invariant mass to the average PDG values~~\cite{PDG}.
A goodness-of-fit
criterion is applied:  $\chi^2/n.d.f. < 50$
 (removing approximately  5\% of signal candidates).

The decays $\Xi_c(2645)^0\to \Xi_c^+\pi^-$ and $\Xi_c(2645)^+\to
\Xi_c^0\pi^+$ are reconstructed by fitting pairs of charged pions and
$\Xi_c$ candidates to a common vertex.
The combinations are accepted if they satisfy the criterion
 $\chi^2/n.d.f. < 10$
 (removing approximately  10\% of signal  candidates)
 and if the momentum of the $\Xi_c\pi$
system in the  
center-of-mass system (CMS) exceeds 2.5~GeV/$c$. Due to the 
hard momentum spectrum of baryons produced in $e^+e^-$ processes, 
this requirement significantly suppresses the combinatorial background.

Figure~\ref{FIG_XC2645} shows a 
 clear $\Xi_c(2645)$ signal in  
$\Xi_c^+\pi^-$ and $\Xi_c^0\pi^+$ mass distributions. 
The second less pronounced maximum above the $\Xi_c(2645)$ peak  is found 
to be a feed-down  of the decay
$\Xi_c(2790)\to \Xi_c^{\prime}(2579)\pi, \Xi_c^{\prime}\to\Xi_c\gamma$ 
(first observed by the CLEO collaboration~\cite{CLEO3}). 
When the photon is missed, the $\Xi_c \pi$ invariant mass
peaks around 2.68 GeV/c$^2$. Both mass and width of the 
feed-down are in agreement with Monte Carlo (MC) expectations.

\begin{figure}[t]
\begin{minipage}[b]{.46\linewidth}
\centering
\setlength{\unitlength}{1mm}
\begin{picture}(95,85)
\put(65,77){\Large\bf (a)}
\put(-5,35){\rotatebox{90}{\large\bf Events / (5~{\rm MeV}/c$^2$)}}
\put(15,-1){{\Large $m(\Xi_c(2645)^0\pi^+)$ [GeV/c$^2$]}}
\includegraphics[height=9.5cm,width=8.5cm]{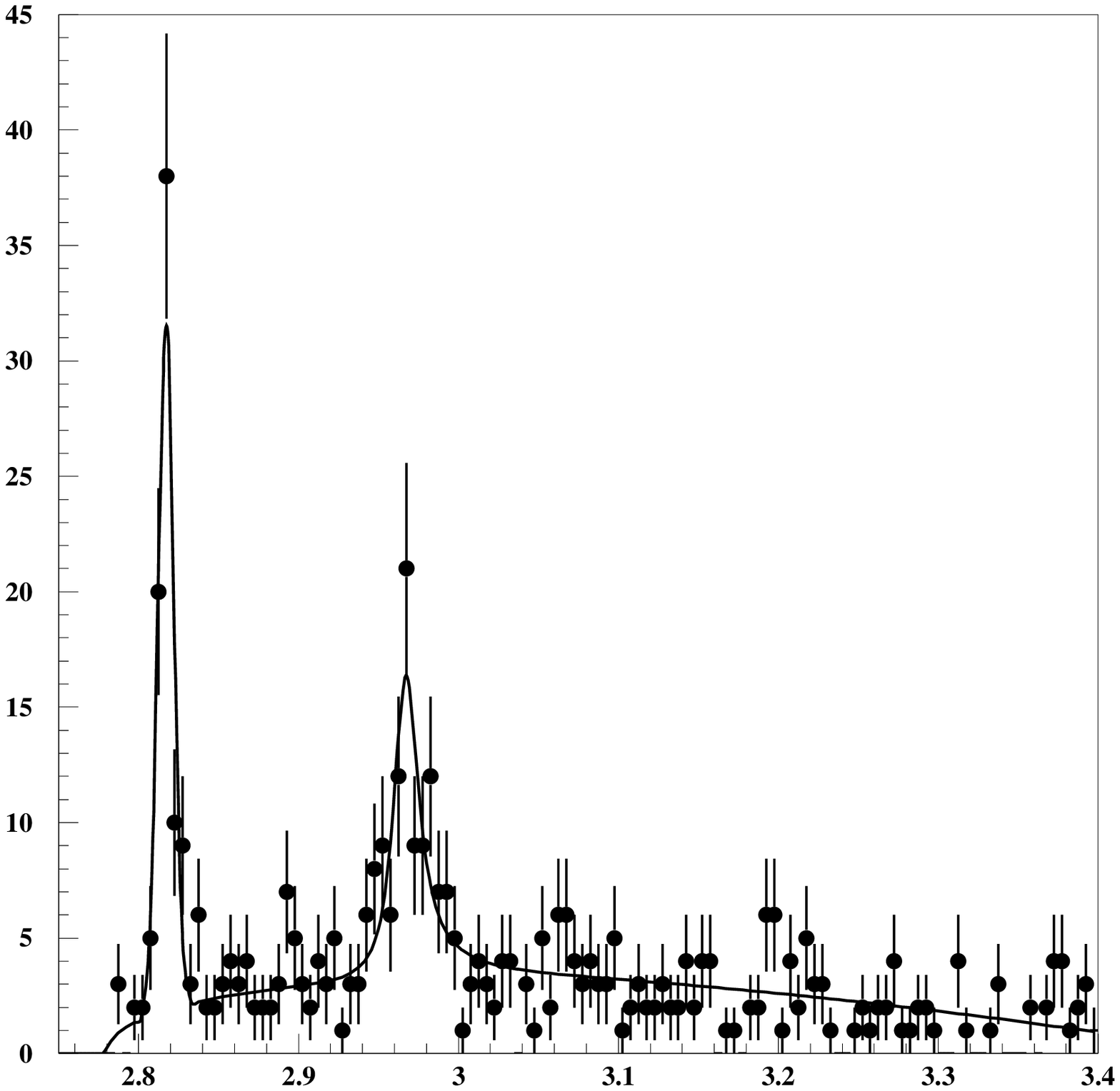}
\end{picture}
\end{minipage}\hfill
\begin{minipage}[b]{.46\linewidth}
\centering
\setlength{\unitlength}{1mm}
\begin{picture}(95,85)
\put(65,77){\Large\bf (b)}
\put(15,-1){{\Large $m(\Xi_c(2645)^+\pi^-)$ [GeV/c$^2$]}}
\put(-5,35){\rotatebox{90}{\large\bf Events / (5~{\rm MeV}/c$^2$)}}
\includegraphics[height=9.5cm,width=8.5cm]{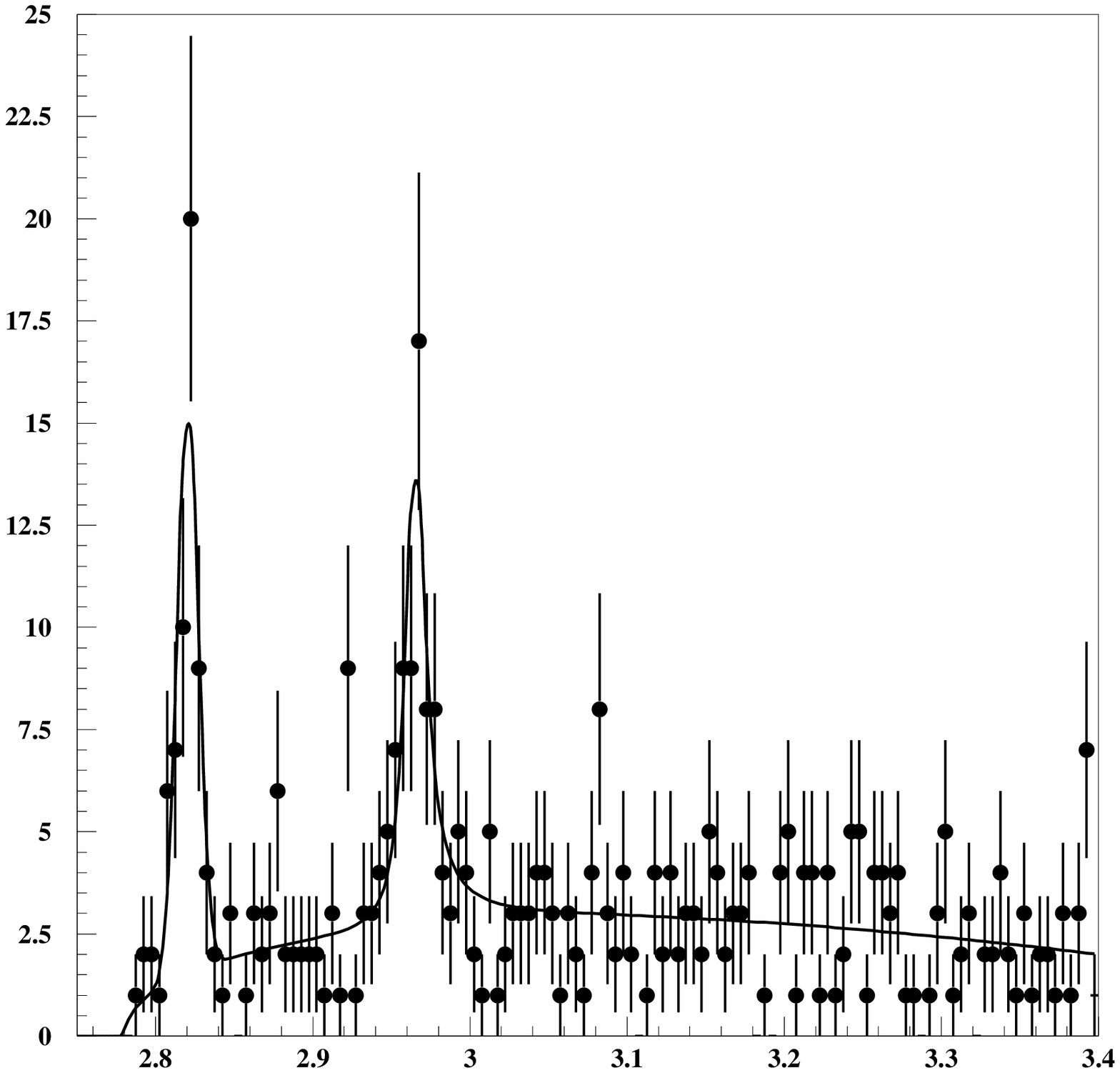}
\end{picture}
\end{minipage}
\caption{Invariant mass distribution for 
{\bf (a)} $\Xi_c(2645)^0\pi^+$ ($\Xi_c(2645)^0\to\Xi_c^+\pi^-$, $\Xi_c^+\to\Xi^-\pi^+\pi^+$) and
{\bf (b)} $\Xi_c(2645)^+\pi^-$ ($\Xi_c(2645)^+\to\Xi_c^0\pi^+$, $\Xi_c^0\to\Xi^-\pi^+$).
Curves correspond to the fit described in the text.} 
\label{FIG_XC2815}
\end{figure}

The decays $\Xi_c(2815)^0\to \Xi_c(2645)^+\pi^-$ and
$\Xi_c(2815)^+\to \Xi_c(2645)^0\pi^+$ are reconstructed by fitting 
the $\Xi_c(2645)$ candidates and an additional charged
pion to a common vertex. Combinations are accepted if they satisfy the criterion
$\chi^2/n.d.f. < 10$
(removing approximately 10\% of signal candidates), and
if the momentum of the $\Xi_c(2645)\pi$ system in the CMS exceeds 2.5
\gev/$c$. The signal region for the $\Xi_c(2645)$ is defined as
(2.635--2.655)~GeV/c$^2$ ($\approx 2.5$ standard deviations) for both decay chains.
Figure~\ref{FIG_XC2815} shows a clear signal of the $\Xi_c(2815)$ baryon
 in the $\Xi_c(2645)^0\pi^+$ and $\Xi_c(2645)^+\pi^-$
mass distributions. Here, we also find a broader peak near 
2.98 GeV/c$^2$ in the 
two charge states.



\section{$\Xi_c(2645)$ mass determination}


We extract the signal yield and the $\Xi_c(2645)^{+,0}$ mass and width
from a fit to the invariant mass distribution of $\Xi_c^{0,+}\pi^{+,-}$ pairs, respectively.
 We use two Gaussians  with a common mean for the signal of the 
$\Xi_c(2645)$:
\begin{equation}
{\cal P}_s(m;\mu,\sigma_1,\sigma_2,f_1)=f_1{\cal G}(m;\mu,\sigma_1)+
(1-f_1){\cal G}(m;\mu,\sigma_2),
\end{equation}
(where the parameter $f_1$ denotes a fractional yield of the first Gaussian)
 and  a single Gaussian ${\cal G}_f$($m;\mu_f,\sigma_f$)  for the feed-down due to 
the $\Xi_c(2790)$:
\begin{equation}
{\cal P}_f(m;\mu_f,\sigma_f)={\cal G}(m;\mu_f,\sigma_f).
\end{equation}

The background 
is described by a threshold function  ($\sqrt{m-m_0}$, where $m_0$ 
corresponds to the threshold mass value)
 multiplied by a fourth-order polynomial $p_4$ with coefficients $c_i, i = 0,1,\ldots 4$:
\begin{equation}
{\cal P}_b(m;m_0,c_0,c_1,c_2,c_3,c_4)=\sqrt{m-m_0} \cdot p_4.
\end{equation}

An additional contribution is due to the reflections from the decay chains
$\Xi_c(2815)^{+,0}\to\Xi_c(2645)^{0,+}\pi^{+,-},~~\Xi_c(2645)^{0,+}\to \Xi_c^{0,+}\pi^0$, where the
neutral pion remains undetected,
 close to the mass peak of the
$\Xi_c(2645)$. The shape of this reflection in $\Xi_c^0\pi^+$ pairs is taken into account 
by fitting  the mass spectra of $\Xi_c^+\pi^+$
in the $\Xi_c(2815)^+\to \Xi_c(2645)^0\pi^+ \to  (\Xi_c^+\pi^-)\pi^+$ decay chain.
Similarly, for the right-sign combinations  $\Xi_c^+\pi^-$, the  invariant mass of the 
wrong-sign pairs  $\Xi_c^0\pi^-$
from the decay chain
 $\Xi_c(2815)^0\to \Xi_c(2645)^+\pi^- \to  (\Xi_c^0\pi^+)\pi^-$
is used. 
The reflection peak is parameterized by a single Gaussian:
\begin{equation}
{\cal P}_r(m;\mu_r,\sigma_r)={\cal G}(m;\mu_r,\sigma_r).
\end{equation}
Thus the overall fit parameterization reads
\begin{equation}
{\cal P} = c_s {\cal P}_s + c_f {\cal P}_f + c_b {\cal P}_b + c_r {\cal P}_r,
\end{equation}
where the yields $c_s$, $c_f$ and $c_b$ are to be determined from the fit.
The yields of the peaks due to the reflections ($c_r$) are estimated according to the formulae
\begin{equation}
c_r = N(\Xi_c(2815)^0)\times \frac{1}{2}\times \frac{\epsilon(\Xi_c^+)}{\epsilon(\Xi_c^0\pi^+)} 
\end{equation}
and
\begin{equation}
c_r = N(\Xi_c(2815)^+)\times \frac{1}{2}\times \frac{\epsilon(\Xi_c^0)}{\epsilon(\Xi_c^+\pi^-)} 
\end{equation}
for $\Xi_c^+\pi^-$ ($\Xi_c^0\pi^+$) pairs, respectively~\cite{SATELIT}.
Here the values of $N(\Xi_c(2815)^{0,+})$ are taken from  Table~\ref{XC285RES} (the results of the fit
to the $\Xi_c(2645)\pi$ mass distribution, described below),
$1/2$ is the isospin factor weight of  $\Xi_c(2645)$ decays, with a $\pi^0$,
to those involving a $\pi^{\pm}$. The efficiencies $\epsilon(\Xi_c^+)=4.55\pm 0.07$\% 
($\epsilon(\Xi_c^0)=7.13\pm 0.14 \%) $
 correspond to the exclusive decays $\Xi_c^+\to \Xi^-\pi^+\pi^+$ ($\Xi_c^0\to \Xi^-\pi^+$), respectively.
They are estimated in our previous measurement (see Table 1 of~\cite{XC1}).
Other parameters determined by the fit are $\mu$,
$\sigma_1$, $\sigma_2$, $f_1$, $\mu_f$ and  $\sigma_f$.

The shape of the background function 
 is fixed from the fit to  the spectrum of $\Xi_c\pi$ invariant masses  
using the  $\Xi_c$ candidates mass sideband: (2.37--2.41) GeV/c/$^2$
 and (2.52--2.57) GeV/c$^2$. 
Results of the fits  are summarized  in Table~\ref{XC264RES}.

\begin{table}[bt]
\begin{center}
\caption{Signal yields and $\Xi_c(2645)$ masses and widths, obtained from the
fits to the $\Xi_c\pi$ mass spectra. $f_1$ denotes the fraction of the
first (narrower) Gaussian; $\sigma_1$ and $\sigma_2$ are the Gaussian widths. Errors shown for
signal yields, $f_1$, $\sigma_1$ and $\sigma_2$ are statistical only.}
\vspace*{0.5ex}
\begin{tabular}{lcccccc}
\hline
Particle	&  \# of  events & ~~~Mass [MeV/c$^2$]~~ & ~~$f_1$~~ & 
~~$\sigma_1$ [MeV]~~  & ~~$\sigma_2$ [MeV]~~ & $\chi^2/n.d.f.$\\ 
\hline
$\Xi_c(2645)^0$  & $611\pm 32$ & $2645.7 \pm 0.2^{+0.6}_{-0.7}$ & $0.44\pm 0.10$ & $1.5\pm 0.3$  & $4.7\pm 0.6$ & 1.69 \\
$\Xi_c(2645)^+$	      & $578\pm 32$ & $2645.6 \pm 0.2^{+0.6}_{-0.8}$ & $0.59\pm 0.11$ & $1.9\pm 0.3$  & $5.5\pm 1.1$ & 1.16 \\
\hline
\end{tabular}
\label{XC264RES}
\end{center}
\end{table}


As a cross-check, the same selection criteria as described above are
also applied to MC samples: $e^+e^-\to c\bar{c}$ and $e^+e^-\to
q\bar{q}$, $q=u,d,s$ with no signal decays included. The background
shapes in the $\Xi_c\pi$ mass spectra for data and MC are in good 
agreement.
The mass of each~$\Xi_c(2645)$ state is obtained from a signal MC sample in
which one $\Xi_c$ decay occurs per event: both are found to be within 
0.2~MeV/c$^2$ of the generated value.

\begin{table}[hbt]
\begin{center}
\caption{Systematic uncertainties on the mass determination of 
the $\Xi_c(2645)$ and $\Xi_c(2815)$.}
\vspace*{0.5ex}
\begin{tabular}{lcccc}
\hline
\multicolumn{1}{c}{Source} &  
\multicolumn{4}{c}{Systematic error [MeV/c$^2$]} \\
\cline{2-5} 
\multicolumn{1}{c}{} & 
\multicolumn{1}{c}{$\Xi_c(2645)^0$} & 
\multicolumn{1}{c}{$\Xi_c(2645)^+$} &
\multicolumn{1}{c}{$\Xi_c(2815)^+$} & 
\multicolumn{1}{c}{$\Xi_c(2815)^0$} \\ 
\hline
(1) Signal width                & 0.1 & 0.1 & 0.3 & 0.5 \\
(2) Fit range                   & 0.0 & 0.0 & 0.1 & 0.0 \\
(3) Bin width                   & 0.1 & 0.0 & 0.1 & 0.4 \\
(4) Background parameterization & 0.1 & 0.0 & 0.0 & 0.2 \\
\hline
(5) Decay length of the $\Xi^-(\Lambda)$  & $^{+0.24}_{-0.30}$ & $^{+0.24}_{-0.30}$
& $^{+0.24}_{-0.30}$ & $^{+0.24}_{-0.30}$ \\
(6) Momentum of the $\Xi^-(\Lambda)$      & $^{+0.25}_{-0.27}$ & $^{+0.24}_{-0.30}$
& $^{+0.25}_{-0.27}$ & $^{+0.24}_{-0.30}$ \\
\hline
(7) Comparison to~\cite{BABARLC}                  & $-0.28$ & $-0.28$ & $-0.28$ & $-0.28$ \\
(8) Azimuthal angle dependence   & $^{+0.17}_{-0.19}$ & $^{+0.17}_{-0.19}$   
& $^{+0.17}_{-0.19}$ & $^{+0.17}_{-0.19}$ \\
(9) CMS momentum  $p^*(\Xi_c(2645))$  dependence  & 0.09 & 0.09 & 0.09 & 0.09 \\
(10) Reflection from the $\Xi_c(2815)$  & $^{+0.1}_{-0.2}$ & $^{+0.1}_{-0.2}$ & n.a. & n.a.  \\
(11) Mass-constrained fit of the $\Xi_c$            & 0.4 & 0.4 & 0.4 & 0.4 \\
\hline
Total systematic error  & $^{+0.6}_{-0.7}$ &$^{+0.6}_{-0.8}$ & $^{+0.7}_{-0.8}$ &$^{+0.9}_{-1.0}$   \\
\hline
\end{tabular}
\label{SYST2645}
\end{center}
\end{table}


The systematic uncertainty on the $\Xi_c(2645)$ mass determination is 
evaluated as follows (Table~\ref{SYST2645}).
First, we consider systematic uncertainties related to the fit procedure.
To take into account imperfect understanding of the signal resolution, we perform 
fits varying the signal widths by their statistical erors,
and compare with values where the widths are floated: the mass 
changes by 0.1 MeV/c$^2$.
For each mode we modify 
 the mass range covered by the fit (extending it by 20\%),
the bin width (2.5--1.0 MeV/c$^2$)
and the parameterization of the background (by varying values of
parameters obtained from the $\Xi_c$ sidebands by $\pm 1~\sigma$).
The resulting changes in the fitted masses are at most  0.1  MeV/c$^2$, depending on the decay.

To estimate the possible dependence of the $\Xi_c(2645)$ mass on the momentum 
and decay length of the $\Xi^-$ and $\Lambda$ hyperons 
we study the decay $\Lambda_c\to \Xi^- K^+ \pi^+$.
A fit to the $\Xi K \pi$ invariant mass distribution
 yields $m(\Lambda_c) = 2286.63\pm 0.09$ MeV/c$^2$ (statistical error only).
The $\Lambda_c$ mass is also determined in bins of the momentum and decay length
of the hyperon $\Xi$, which leads to systematic uncertainties of
$^{+0.25}_{-0.27}$ MeV/c$^2$ and
$^{+0.24}_{-0.30}$ MeV/c$^2$, respectively.

To test the modeling of the detector response (alignment, uniformity of
magnetic field, correct treatment of specific ionization and scattering
in the material), which could cause a bias in the overall mass scale, we study 
 $\Lambda_c^+\to p K \pi$
decays. 
A fit to the $p K \pi$ invariant mass distribution 
yields $ m(\Lambda_c) = 2286.74\pm 0.02$ MeV/c$^2$ 
(statistical error only).
The above value is compared to the recent  measurement 
by the BaBar collaboration~\cite{BABARLC}, which yields
$ m(\Lambda_c) = 2286.46\pm 0.14$ MeV/c$^2$.
As a result, a $-0.28$ MeV/c$^2$ shift is assigned as a systematic error.
The mass of the $\Lambda_c$ reconstructed in $p K\pi$ is also determined
in bins of the azimuthal angle. The maximal deviations with respect to the
value given above are assigned as the corresponding systematic errors,  
yielding $^{+0.17}_{-0.19}$ MeV/c$^2$. 
The same study, performed in bins of $\Lambda_c$ center-of-mass
momentum provides an estimate of $\pm 0.09$ MeV/c$^2$ as the
respective systematic uncertainty.

The uncertainty on the parameters of the reflection due to the decays
of $\Xi_c(2815)$ results in a systematic error of $^{+0.1}_{-0.2}$ MeV/c$^2$
estimated by performing the fit with the removal of 
the reflection contribution and also by varying its width and yield within
their statistical errors.

The $\Xi_c(2645)$ mass also depends on the
value of m($\Xi_c$) applied in the mass-constrained fit.
A change of  m($\Xi_c$) almost linearly 
transforms to a shift in the measured value of m($\Xi_c(2645)$).
As a result, we include a systematic uncertainty
equal to the statistical error in the determination of the $\Xi_c$
mass~\cite{PDG},
 i.e.  $\pm 0.4$ MeV/c$^2$ both for the $\Xi_c^+$ and
 $\Xi_c^0$. 


It is also checked that the measured mass value   
is  stable within one standard deviation while fitting separately the spectra
corresponding to particles and antiparticles in the final state. 
The total systematic uncertainty is obtained by adding the individual
 contributions in quadrature.

%

The masses of the $\Xi_c(2645)^+$ and $\Xi_c(2645)^0$ (Table~\ref{XC264RES}) are
 in agreement with, and more accurate than the current PDG averages~\cite{PDG}.
Assuming that uncertainties (5)--(10) from Table~\ref{SYST2645}
are the same for charged and neutral $\Xi_c(2645)$'s and, as such,
cancel in the $\Xi_c(2645)^+ - \Xi_c(2645)^0$ mass splitting,
 we find the mass difference between the charged and neutral states to be:
\begin{equation} 
m_{\Xi_c(2645)^+} - m_{\Xi_c(2645)^0} = (-0.1 \pm 0.3 ({\rm stat}) \pm 0.6 {(\rm syst}))~{\rm MeV/c}^2.
\label{MASSSPL}
\end{equation}


\section{$\Xi_c(2815)$ mass determination}


For each decay mode, we extract the signal yield and the $\Xi_c(2815)$ 
mass and width
from a fit to the invariant mass distribution of $\Xi_c(2645)\pi$ pairs.
We use a  single Gaussian for the signal from 
the $\Xi_c(2815)$ and a Breit-Wigner shape convoluted with a Gaussian for the peak 
near $m(\Xi_c(2645)\pi)= 2980$~MeV/c$^2$ (we denote this peak 
as $\Xi_c(2980)$). The width of the latter Gaussian, describing the 
experimental mass resolution, is fixed from the MC simulation
to the value 
of $2.2\pm 0.2 $ MeV.
The background is parameterized by a threshold function multiplied by a  first-order polynomial with
coefficients $d_0$ and $d_1$:
\begin{equation}
{\cal P}_b(m;m_0,d_0,d_1)= {\rm atan} (\sqrt{m-m_0})\times p_1.
\end{equation}
 The fit results are summarized in 
Tables~\ref{XC285RES} and~\ref{XC298RES} for the
$\Xi_c(2815)$ and $\Xi_c(2980)$ signals, respectively.

\begin{table}[htb]
\begin{center}
\caption{Signal yields and  $\Xi_c(2815)$ masses and widths, obtained from the fits to the $\Xi_c(2645)\pi$ mass spectra.} 
\vspace*{0.5ex}
\begin{tabular}{lrccc}
\hline
Particle	&  \# of  events	& Mass [MeV/c$^2$] & Gaussian width [MeV] & $\chi^2/n.d.f.$ \\ 
\hline
$\Xi_c(2815)^+$  & $72.5\pm 9.6$ & \hspace*{0.3cm}$2817.0 \pm 1.2(\rm stat) ^{+0.7}_{-0.8}(\rm syst)$ & $4.9\pm 0.9$ & 1.03\\
\hline
$\Xi_c(2815)^0$	      & $47.5\pm 7.8$ & \hspace*{0.3cm}$2820.4 \pm 1.4(\rm stat) ^{+0.9}_{-1.0}(\rm syst)$ & $6.9\pm 1.1$ & 0.97\\
\hline
\end{tabular}
\label{XC285RES}
\end{center}
\begin{center}
\caption{Signal yields and $\Xi_c(2980)$ masses and natural widths, obtained from the fits to the $\Xi_c(2645)\pi$ mass spectra.} 
\vspace*{0.5ex}
\begin{tabular}{lrccc}
\hline
Particle	&  \# of  events	& Mass [MeV/c$^2$] & $\Gamma$, natural width [MeV] & Significance [$\sigma$]\\ 
\hline
$\Xi_c(2980)^+$  &  $78.3\pm 13.4$  & \hspace*{0.3cm}$2967.7 \pm 2.3(\rm stat)^{+1.1}_{-1.2}{\rm (syst)}$ & $18\pm 6\pm 3$ & 7.3 \\
\hline
$\Xi_c(2980)^0$	      &  $56.9\pm 12.5$  & \hspace*{0.3cm}$2965.7 \pm 2.4(\rm stat)^{+1.1}_{-1.2}{\rm (syst)}$ & $15\pm 6\pm 3$ & 6.1 \\
\hline
\end{tabular}
\label{XC298RES}
\end{center}
\end{table}

The systematic uncertainties on the $\Xi_c(2815)$ mass determination 
are estimated following the procedure used for the $\Xi_c(2645)$.
The total systematic uncertainty is obtained by adding the individual
contributions in quadrature (Table~\ref{SYST2645}).

The masses of the $\Xi_c(2815)^+$ and $\Xi_c(2815)^0$ (Table~\ref{XC285RES}) are
 in agreement with the CLEO~\cite{CLEO2} measurements.
For the charged state the accuracy is comparable to~\cite{CLEO2}, while for the
neutral one it is better.
Assuming that uncertainties (5)--(9) from Table~\ref{SYST2645}
are the same for charged and neutral $\Xi_c(2815)$ and as such
cancel in the $\Xi_c(2815)^+ - \Xi_c(2815)^0$ mass splitting,
 we find the mass difference between the charged and neutral states to be
\begin{equation} 
m_{\Xi_c(2815)^+} - m_{\Xi_c(2815)^0} =  (-3.4 \pm 1.9 ({\rm stat}) \pm 0.9 {(\rm syst}))~{\rm MeV/c}^2.
\label{DMX298}
\end{equation} 


\section{Observation of $\Xi_c(2980)\to \Xi_c(2645)\pi$}


The mass of the  $\Xi_c(2980)^0$ (Table~\ref{XC298RES}) is compatible
 with the masses of the $\Xi_c(2980)^0$, decaying to 
$\Lambda_c^+ K^0_s \pi^-$, as observed by Belle~\cite{RUSLAN} and confirmed by BaBar~\cite{BABARXC}.
 The width of the $\Xi_c(2980)^0$ is smaller, but statistically consistent  with the value
measured by  BaBar:~($31\pm 7\pm 8$)~MeV. 

For the charged state $\Xi_c(2980)^+$, the mass given in Table~\ref{XC298RES}
is in agreement with the value determined by BaBar (($2969.3\pm 2.2\pm 1.7$)~MeV/c$^2$) and 
smaller than the result in our observation of 
$\Xi_c(2980)\to \Lambda_c^+ K^- \pi^+$ 
($2978.5\pm 2.1\pm 2.0$)~MeV/c$^2$. 
The fitted width of the $\Xi_c(2980)^+$ is smaller than  the value found by Belle ($43.5\pm 7.5\pm 7.0$)~MeV 
and BaBar~($27\pm 8\pm 2$)~MeV.
To estimate the significance of the $\Xi_c(2980)^{+,0}$ observation,
the fit is repeated omitting the signal component due to this state from the fit. The significance
is determined from $-2\ln{({\cal L}_0/{\cal L})}$, where ${\cal L}$ and ${\cal L}_0$ refer to the
maximum of the default likelihood function (describing also the $\Xi_c(2980)$) and the likelihood
function omitting this signal component, respectively. This quantity should be distributed 
as $\chi^2({\mathrm n.d.f.} = 3)$, as three parameters are free for the signal.

The systematic uncertainties on the $\Xi_c(2980)$ mass  
are determined following the procedure used for the $\Xi_c(2645)$ and $\Xi_c(2815)$.
 The uncertainty due to the fit procedure
is determined  to be $\pm 0.9$~MeV/c$^2$ by varying the bin width and the experimental resolution 
within its error ($2.2\pm 0.2$)~MeV and by 
fitting the background with a  second-order polynomial. The above mentioned procedures 
of varying the bin width and the experimental resolution within its error  are also used
to determine the systematic uncertainty of the natural width 
of the $\Xi_c(2980)$ (Table~\ref{XC298RES}).
The total systematic uncertainty on the $\Xi_c(2980)$ mass determination
(Table~\ref{XC298RES}) is obtained by adding 
in quadrature the uncertainty due to the fit procedure and the
contributions (5--9) and (11), as given in  Table~\ref{SYST2645}.

Given the uncertainties in the measured masses and widths of 
the $\Xi_c(2980)$, this state is consistent with the
charmed baryon observed in $\Lambda_c K \pi$ final state, the $\Xi_c(2980)$. 
No signals are observed in the $\Xi_c(2645)\pi$ mass spectra near 
the masses of 
3055, 3077 and 3123~MeV/c$^2$, corresponding to the new states  observed by 
Belle~\cite{RUSLAN} and BaBar~\cite{BABARXC} in $\Lambda_c K \pi$ decays.




\section{Conclusions}

Based on a large sample of the $\Xi_c$ hyperons,
 the masses of $\Xi_c(2645)$ and $\Xi_c(2815)$ baryons  are measured (Table~\ref{PDGTABLE})
together with the mass splittings within isospin doublets:
$$m_{\Xi_c(2645)^+} - m_{\Xi_c(2645)^0} = (-0.1 \pm 0.3 ({\rm stat}) \pm 0.6 {(\rm syst}))~{\rm MeV}/c^2,$$
$$m_{\Xi_c(2815)^+} - m_{\Xi_c(2815)^0} =  (-3.4 \pm 1.9 ({\rm stat}) \pm 0.9 {(\rm syst}))~{\rm MeV/c}^2,$$
They are determined
with a much better precision than the current world averages. The measurement also provides
 the first confirmation
of the respective CLEO observations \cite{CLEO1,CLEO2}.

\begin{table}[h]
\begin{center}
\caption{Masses of the $\Xi_c(2645)$ and $\Xi_c(2815)$.}
\vspace*{0.5ex}
\begin{tabular}{ccc}
\hline
\multicolumn{1}{c}{Particle}     & 
\multicolumn{2}{c}{Mass [MeV/c$^2$]}            \\ 
\cline{2-3}
\multicolumn{1}{c}{}                            & 
\multicolumn{1}{c}{PDG}                         &
\multicolumn{1}{c}{This study }                 \\
\hline
$\Xi_c(2645)^+$   &   ~~~$2646.6\pm 1.4$~~~  & $2645.6\pm 0.2 {\rm (stat)} ^{+0.6}_{-0.8} {\rm (syst)}$ \\
$\Xi_c(2645)^0$   &   ~~~$2646.1\pm 1.2$~~~  & $2645.7\pm 0.2 {\rm (stat)} ^{+0.6}_{-0.7} {\rm (syst)}$ \\
$\Xi_c(2815)^+$   &   ~~~$2816.5\pm 1.2$~~~  & $2817.0\pm 1.2 {\rm (stat)} ^{+0.7}_{-0.8} ({\rm syst})$ \\
$\Xi_c(2815)^0$   &   ~~~$2818.2\pm 2.1$~~~  & $2820.4\pm 1.4 {\rm (stat)} ^{+0.9}_{-1.0} ({\rm syst})$ \\
\hline
\end{tabular}
\label{PDGTABLE}
\end{center}
\end{table}

In the $\Xi_c(2645)^+\pi^-$ and $\Xi_c(2645)^0\pi^+$ spectra, two states
with masses around 2980~MeV/c$^2$ are observed with large statistical significance.
The measured masses and widths of the $\Xi_c(2980)$ are slightly different from the values determined for the
$\Xi_c(2980)$ in previous measurements~\cite{RUSLAN,BABARXC}, 
although still consistent within the uncertainties.
While identification of this state as the $\Xi_c(2980)$ (\cite{RUSLAN,BABARXC}) is plausible,
further  high-statistics measurements of its properties would be welcome
to confirm this hypothesis.


\section{Acknowledgements}


We thank the KEKB group for the excellent operation of the
accelerator, the KEK cryogenics group for the efficient
operation of the solenoid, and the KEK computer group and
the National Institute of Informatics for valuable computing
and Super-SINET network support. We acknowledge support from
the Ministry of Education, Culture, Sports, Science, and
Technology of Japan and the Japan Society for the Promotion
of Science; the Australian Research Council and the
Australian Department of Education, Science and Training;
the National Natural Science Foundation of China under
contract No.~10575109 and 10775142; the Department of
Science and Technology of India; 
the BK21 program of the Ministry of Education of Korea, 
the CHEP SRC program and Basic Research program 
(grant No.~R01-2005-000-10089-0) of the Korea Science and
Engineering Foundation, and the Pure Basic Research Group 
program of the Korea Research Foundation; 
the Polish State Committee for Scientific Research; 
the Ministry of Education and Science of the Russian
Federation and the Russian Federal Agency for Atomic Energy;
the Slovenian Research Agency;  the Swiss
National Science Foundation; the National Science Council
and the Ministry of Education of Taiwan; and the U.S.\
Department of Energy.



\begin{thebibliography}{1}
\bibitem{CHGCONJ} Charge-conjugate modes are implicitly included everywhere,
 unless specified otherwise.
\bibitem{ZAXIS} The $z$-axis is oriented opposite to the 
direction of the $e^+$ beam, along the symmetry axis of the detector.
\bibitem{PROTON} For protons attributed to the decay $\Lambda\to p\pi$,
  the hadron identification criterion was relaxed relative to
  the one described in the section ``Track reconstruction and
  identification''.
\bibitem{SATELIT} In the denominators of (6) and (7) 
we use 
$\epsilon(\Xi_c^{0,+})$ instead of
$\epsilon(\Xi_c^{0,+}\pi^{+,-})$
 as the charged pions are reconstructed and
 identified at Belle with a high overall efficiency, exceeding 80~\%.
As described in the text, this approximation is considered
conservatively in the estimation of the systematic error.
\end{thebibliography}
\end{document}